\documentclass[preprint2]{aastex}
\usepackage{natbib,graphicx,xspace,color}

\singlespace

\shorttitle{Weak Lensing of Bright X-Ray Abell Clusters}
\shortauthors{Cypriano et al.}

\newcommand\eq{\begin{equation}}
\newcommand\eeq{\end{equation}}
\newcommand\eqn{\begin{eqnarray}}
\newcommand\eeqn{\end{eqnarray}}
\newbox\grsign \setbox\grsign=\hbox{$>$} \newdimen\grdimen
\grdimen=\ht\grsign
\newbox\simlessbox \newbox\simgreatbox
\setbox\simgreatbox=\hbox{\raise.5ex\hbox{$>$}\llap
     {\lower.5ex\hbox{$\sim$}}}\ht1=\grdimen\dp1=0pt
\setbox\simlessbox=\hbox{\raise.5ex\hbox{$<$}\llap
     {\lower.5ex\hbox{$\sim$}}}\ht2=\grdimen\dp2=0pt
\newcommand\simgreat{\mathrel{\copy\simgreatbox}}

\newbox\simppropto
\setbox\simppropto=\hbox{\raise.5ex\hbox{$\sim$}\llap
     {\lower.5ex\hbox{$\propto$}}}\ht2=\grdimen\dp2=0pt

\def\kms{km s$^{-1}$\xspace} 
\def\ergs{erg s$^{-1}$\xspace}
\def\lx{L$_X$\xspace}
\def\tx{T$_X$\xspace}

\begin{document}

\title{Weak lensing mass distributions for 24 X-Ray Abell Clusters.
\thanks{Based on observations collected with ESO Very Large Telescope
Antu (UT1)}}

\author{Eduardo S. Cypriano and Laerte Sodr\'e Jr.}
\affil{Departamento de Astronomia, Instituto de Astronomia, Geof\'{\i}sica e
Ci\^encias Atmosf\'ericas da USP, Rua do Mat\~ao, 1226, 05508-900,
S\~ao Paulo, Brazil} 
\email{eduardo,laerte@astro.iag.usp.br}

\author{Jean-Paul Kneib}
\affil{Observatoire Midi-Pyr\'en\'ees, Laboratoire d'Astrophysique,
UMR 5572, 14 Avenue E. Belin, 31400 Toulouse, France and
Caltech, Astronomy, 105-24, Pasadena, CA 91125, USA }
\email{kneib@ast.obs-mip.fr,kneib@caltech.edu}

\author{Luis E. Campusano}
\affil{Observatorio Astron\'omico Cerro Cal\'an, Departamento de 
Astronom\'{\i}a, Universidad de Chile, Casilla 36-D, Santiago, Chile}
\email{luis@das.uchile.cl}


\begin{abstract}
We use the weak gravitational lensing effect to study the mass
distribution  and dynamical
state of a sample of 24 X-ray luminous clusters of galaxies ($0.05<z<0.31$)
observed with the FORS1 instrument mounted on the VLT-Antu
(Unit Telescope 1) under homogeneous sky conditions and subarsecond
image quality.  The galaxy shapes were measured in the combined
V, I, R image after deconvolution with a locally determined
point-spread-function, while the two-dimensional mass distributions
of the clusters were computed using an algorithm based on the maximum
entropy method. By comparing the mass and light distributions
of the clusters in our sample, we find that their mass centers, for the majority
of the clusters, is consistent with the positions of optical centers.
We find that some clusters present significant mass substructures which
generally have optical counterparts. 
At least in one cluster (Abell 1451), we detect a mass substructure 
without an obvious luminous counterpart. 
The radial profile of the shear of the clusters was fitted using circular
and elliptical isothermal elliptical  distributions, which
allowed the finding of a strong correlation between the orientation
of the major-axis of the matter distribution and  the corresponding major-axes
of the brightest cluster galaxy light-profiles. 
Estimates of how close to dynamical relaxation are these clusters
were obtained through comparison of our weak-lensing mass measurements
with the x-ray and velocity dispersion determinations available in
the literature. We find that clusters with intra-cluster gas colder than 8 keV
show a good agreement between the different mass determinations, but
clusters with gas hotter than 8 keV present discrepant
mass values. The clusters diagnosed to be out of equilibrium are Abell 1451,
2163 and 2744, all of them having hints of substructure. Abell 2744 presents the
largest discrepancy between its X-ray and weak-lensing temperature determinations,
which can be interpreted as being due to the interaction between the two
kinematical components along the line of sight found by \citet{girardi}.

\end{abstract}


\section{Introduction}

The emergence of mass structures in the universe is currently linked
to primordial density perturbations, dominated by cold dark matter,
through a hierarchical clustering process involving gravitational
instabilities.
A natural consequence of this bottom-up scenario is that
the most massive structures form later in time, and, depending on the
world models, they are expected to show signatures of their assembling history.
Since galaxy clusters are possibly the largest (nearly) virialized 
structures \citep[e.g.][]{P&S}, the study of their main properties (masses, 
shapes, radial profiles etc.) can provide
valuable clues on the details of the agglomeration of matter
in the universe and
hence on the nature of the dark matter \citep[e.g.][]{kauffmann99}.
Clusters can also be used as cosmological probes providing additional data that,
together with cosmic microwave background observations, can be used on the
determination of  several cosmological parameters.
Studies of their mass 
function \citep[e.g.][]{H&A,xmass}, mass-to-light ratio
\citep[e.g.][]{BLD95,CNOC} and baryon fraction  \citep[e.g.][]{WF,allen2002}
can provide constraints on key cosmological parameters, like the mass density
parameter, $\Omega_M$, the bias,  and on the power-spectrum amplitude and shape
parameter \citep{l&c94,B&F98}. 

Galaxy clusters are complex systems that hold together galaxies, hot gas
and dark matter. These components are governed by different physical
mechanisms and their study requires the use of different observational
techniques. Imaging in visible light, reveals the cluster through
their member galaxies which we know now that contribute only a
small fraction of the cluster total mass. But, if these galaxies
are in virial equilibrium, the depth of the cluster potential well
can be accessed through their velocity dispersion, although
the continuous accretion of field galaxies can bias the
mass measurements to higher values \citep[e.g.][]{laerte89}. 
The thin hot gas-- mainly hydrogen--  that permeates the cluster
gravitational potential is found to be at temperatures of the order
of $10^7-10^8$ K, thus fully ionized and emitting X-rays via
thermal {\it bremsstrahlung} process \citep[see][for a more complete discussion]
{Sarazin}. The emission in the X-ray  band provides an efficient
method to find galaxy clusters, and estimates of the cluster masses
if the hypothesis that the gas is in hydrodynamical equilibrium
is adopted. The advantage of using the hot intracluster gas rather than
the member galaxies for mass determinations is that it has a much
shorter relaxation time due to its self-interactivity \citep[][]{Sarazin}.
Nevertheless, there is evidence, that many clusters present
significant departures from dynamical relaxation \citep[e.g.][]{GB,laertinho}.
Presumably the clusters that are in the top or the mass function will
achieve last their formation process, so we expect to find some degree
of correlation between the departures from equilibrium and the mass
of the clusters.

The development of gravitational lensing techniques, for both the strong 
and weak regimes, has presented a new way of measuring masses regardless 
the nature or dynamical state of the matter \citep{F&M94,mellier99}.
Strong lensing, that relies on the modeling of systems of
multiply lensed background galaxies, has proved to be an accurate method,
mainly when the number of gravitational arcs is large \citep[e.g.][]{kneib96}.
However, this technique can probe only the very central regions of the
clusters, and is limited to the most massive and concentrated objects.
On the other hand, the statistical analysis of the weak distortion caused 
by the cluster shear field on images of faint background galaxies allows the
mapping of the matter distribution to  much larger radius; besides, this
gravitational distortion is detectable in almost all clusters when
large telescopes with current instrumentation are used.
After the seminal work of \citet{tyson}, the weak-lensing technique started
to be used for the study of some clusters having special interest because of
extreme values in some of their measured properties, and new methodologies
were established
\citep[see][for a review]{mellier99}. 

With the development of wide field cameras and the increase in the
number of large telescopes, it is now possible 
to study large samples of galaxy clusters selected using well defined
criteria, ideally observed under similar conditions, 
and analyzed with the same technique. 
\citet{dahle} studied 38 Northern
X-ray luminous clusters using data collected with 2-meter class telescopes and
selected to be representative of the most massive clusters in the
$0.15<z<0.35$ redshift range. However, the clusters in this sample 
were not all observed with the same instruments and, sometime, with varying 
atmospheric conditions. In this paper
we present a weak-lensing analysis of a sample of
24 X-ray luminous southern clusters with $0.05<z<0.31$, based on
imaging observations with the ESO-VLT telescope taken under
sub-arcsecond conditions. The clusters were selected from the
X-ray Bright Abell Cluster Survey \citep[XBACS][]{XBACS},
with $-50\degr< \delta < 15\degr$, 12 h $\le \alpha <$ 1 h,
$L_X > 5\times 10^{44}$ erg s$^{-1}$. Similar
observations have been done of a 
complementary sample of 27 X-ray clusters, with 1 h $\le \alpha <$ 12 h , whose
weak-lensing analysis is underway. Hereafter, and due to the rather rare 
choice of a lower-z limit of 0.05, we will refer to our
whole VLT survey as the Low-z Lensing Survey of X-ray Luminous Clusters
(LZLS), and the sample presented here is its Part I. The data used
in this paper derive from observations primarily designed for detection of
strong-lensing features but conceived for weak-lensing measurements
also; the service observing mode at ESO was essential to insure
the high quality and homogeneity of the images. The statistics of the
occurrence of bright arcs for the whole LZLS survey will be 
presented in a separate paper \citep[][]{paper2}, together with its
implications with respect to the clusters inner mass radial profile.
In future papers, 
we will address the combined evaluation of the mass distribution
we will use both weak and strong lensing in those LZLS clusters
with gravitational arcs focusing on those involving arcs with redshift
determinations.

In this paper we determine the mass distributions
for the galaxy clusters in our sample and their total masses using weak-lensing
techniques, and investigate their dynamical state 
through comparison of the weak-lensing masses  with already published
virial and X-ray mass estimates. In Section 2 we describe the sample
selection and the observations. In Section 3 we present the procedures
adopted in the weak-lensing analysis, including galaxy shape measurements
and the reconstruction of cluster density maps, as well as the results
obtained. The discussion of the results is presented in Section 4
together with the comparison with dynamical, X-ray and weak-lensing
masses taken from the literature. In Section 5, we summarize
our main conclusions.  In the Appendix, we display images, mass and light maps,
and weak-shear profiles for the cluster sample.

Throughout this paper we adopt $\Omega_M$ = 0.3, $\Omega_\Lambda$ = 0.7 and
H$_0$ = 70 \kms Mpc$^{-1}$.

\section{Observations \label{obs}}

\subsection{Sample Selection}

The galaxy clusters investigated in this work were selected based on
their high X-ray luminosities (\lx). Initially, we selected  clusters in
the  X-Ray Brightest Abell Cluster Catalog (XBAC) \citep{XBACS} with
\lx $\ge 5 \times 10^{44}$\ergs in the 0.1-2.4 keV band, which constitutes
a good threshold to identify real, massive clusters \citep[e.g.][]{luppino}.
We then restricted the sample to clusters that could be reached with
ESO VLT ($-50\degr \le \delta < +15\degr$).
A low-z limit of 0.05 was chosen, because for lower
redshifts the angular size of the clusters is too large compared
with the field of the imaging camera. For $z<0.05$, possible
gravitational arcs would appear projected deep inside the central galaxy image
and the angular size of cluster galaxies in the camera field would be very high,
making hard to find a suitable number of background galaxies
(without strong light contamination) for a weak-lensing analysis.
ESO VLT was considered a very good choice because the observations
could be completed with some 10 hours of service observing and the
homogeneity and image quality of the observations could be optimally
achieved. These criteria produced 27 targets within 12 h $\le \alpha <$ 1 h.
Three of these clusters (A68, A1689 and A1835)
have been already observed with HST and CFHT by one of us (JPK)
and the results will be published elsewhere \citep{graham,sebastien}.
The remaining 24 rich galaxy clusters that are studied in this paper
are presented in  Table \ref{VMACS}. This sample constitutes the
Part I of the VLT Low-z Lensing Survey of X-ray Luminous Clusters
(LZLS).

\begin{deluxetable}{lrrcccc}
\tablewidth{0pt}
\tablecaption{VLT Massive Abell Cluster Sample \label{VMACS}}
\tablehead{
\colhead{(1)} & \colhead{(2)} & \colhead{(3)} &
\colhead{(4)} & \colhead{(5)} & \colhead{(6)} & \colhead{(7)}\\
\colhead{Name} & \colhead{RA} & \colhead{DEC} &
\colhead{z} & \colhead{L$_X$} & \colhead{seeing}&
\colhead{$\mu_{sky}$ (1 $\sigma$)} \\
\colhead{} & \colhead{} & \colhead{} &
\colhead{} & \colhead{(10$^{44}$ \ergs)} & \colhead{(arcsec)} &
\colhead{(R mag arcsec$^{-2}$)}}
\startdata
A2744$^*$ & ~0 14 16.1 &   -30 22 58.8 & 0.308 & 22.05  & 0.59 & 26.92 \\
A22       & ~0 20 38.6 &   -25 43 19.2 & 0.131 & ~5.31  & 0.60 & 26.71 \\
A85       & ~0 41 48.7 &   ~-9 19 04.8 & 0.052 & ~8.38  & 0.59 & 26.58 \\
A2811     & ~0 42 07.9 &   -28 32 09.6 & 0.108 & ~5.43  & 0.64 & 26.77 \\
A1437     & 12 00 25.4 &~\phs3 21 03.6 & 0.134 & ~7.72  & 0.52 & 26.22 \\
A1451     & 12 03 14.6 &   -21 31 37.2 & 0.171 & ~7.40  & 0.48 & 26.37 \\
A1553     & 12 30 48.0 &\phs10 33 21.6 & 0.165 & ~7.05  & 0.80 & 26.10 \\
A1650     & 12 58 41.8 &   ~-1 45 21.6 & 0.084 & ~7.81  & 0.68 & 26.28 \\
A1651     & 12 59 24.0 &   ~-4 11 20.4 & 0.085 & ~8.25  & 0.76 & 26.31 \\
A1664     & 13 03 44.2 &   -24 15 21.6 & 0.128 & ~5.36  & 0.79 & 26.51 \\
A2029     & 15 10 55.0 &~\phs5 43 12.0 & 0.077 & 15.35  & 0.42 & 26.21 \\
A2104     & 15 40 06.5 &   ~-3 18 21.6 & 0.155 & ~7.89  & 0.41 & 26.30 \\
A2163     & 16 15 49.4 &   ~-6 09 00.0 & 0.208 & 37.50  & 0.43 & 26.43 \\
A2204     & 16 32 46.8 &~\phs5 34 26.4 & 0.152 & 20.58  & 0.49 & 26.25 \\
A3695     & 20 34 46.6 &   -35 49 48.0 & 0.089 & ~5.07  & 0.76 & 26.63 \\
A3739     & 21 04 17.5 &   -41 20 20.4 & 0.179 & ~7.00  & 0.70 & 26.66 \\
A2345     & 21 26 58.6 &   -12 08 27.6 & 0.176 & ~9.93  & 0.56 & 26.37 \\
A2384     & 21 52 16.6 &   -19 36 00.0 & 0.094 & ~6.82  & 0.61 & 26.28 \\
A2426     & 22 14 32.4 &   -10 21 54.0 & 0.099 & ~5.10  & 0.81 & 26.64 \\
A3856     & 22 18 37.4 &   -38 53 13.2 & 0.126 & ~6.40  & 0.77 & 26.85 \\
A3888     & 22 34 32.9 &   -37 43 58.8 & 0.151 & 14.52  & 1.03 & 26.75 \\
A3984     & 23 15 37.7 &   -37 44 52.8 & 0.178 & ~9.18  & 0.53 & 26.92 \\
A2597     & 23 25 16.6 &   -12 07 26.4 & 0.085 & ~7.97  & 0.52 & 26.64 \\
A4010     & 23 31 14.2 &   -36 30 07.2 & 0.096 & ~5.55  & 0.55 & 26.92 \\
\enddata
\tablecomments{(1) Cluster name. (2),(3) Equatorial J2000 coordinates. (4)
Redshift.
(5) X-ray luminosities in the [0.1-2.4] keV band \citep{XBACS}.
(6) FWHM of stellar images in
the combined VRI image. (7) Surface brightness corresponding to 1$\sigma$ 
above the sky average in the R image. ($*$) A2744 is also known as AC118.}
\end{deluxetable}

\subsection{Imaging and Data Reduction}

All clusters were observed in service mode using VLT Antu (Unit Telescope 1),
from April to July 2001 (ESO program 67.A-0597), with the
FORS1 instrument (FOcal Reducer/low
dispersion Spectrograph) working in its imaging mode. FORS1 employs a
TK2048EB4-1 backside illuminated thinned CCD, with $2048\times 2048$
pixels, each with  $24\times 24 \mu^2$ area. The employed
imaging mode of FORS1 results in 
a pixel scale of 0\arcsec.2 and a square field of view of 6\arcmin.8 on a side,
corresponding to
0.4, 0.9 and 1.8 Mpc at the smaller, median and greater redshifts of the clusters
of our sample, which are 0.05, 0.126 and 0.305 respectively.
The imaging was chosen to be centered on the clusters cores and done
through the V, R, and I-bands with exposure times of 330s in each filter,
ensuring a good detection of faint galaxies in all bands. The color
information is essential to discriminate elongated objects belonging
to the cluster from the background sources needed for the weak-lensing
measurements. All the observations were conducted under good sky transparency and
and an excellent image quality characterized by a median seeing of  0\arcsec.6,
ranging from 0\arcsec.4 to 1\arcsec.0.

The data reduction were done in a standard way
using {\sc IRAF} packages. For each cluster we combined the images in the 
three filters, normalizing by their modes, to produce a high signal-to-noise 
``VRI'' image that was used for the weak-lensing analysis.
The images in each
band were used for the photometry of the objects in each cluster field.
Comparing stellar shapes measured in the VRI images with the
single-band ones, we concluded that the combination process did not
introduce biases. Tests on galaxies revealed that the use of the combined VRI
image improves the precision in the shape measurements in comparison to
single-band measurements.

Zero-point magnitudes, extinction, and color coefficients for transformations
from instrumental to standard magnitudes were provided by the ESO pipeline.
Almost all data have been collected under photometric sky conditions 
or nearly so.
Table  \ref{VMACS} contains, besides general informations about each cluster
in our sample, the seeing (FWHM) derived from stellar images in the combined 
VRI image. The table also includes the surface brightness corresponding to 
a 1$\sigma$ fluctuation above the sky background in the R image,
which is a measure of the quality and depth of these images.

\section{Weak-Lensing Analysis}

\subsection{Construction of the Catalog}

The detection of objects in the images and the extraction of their 
main parameters were made with {\sc SExtractor} \citep{sex}. We run this 
program separately for the single band images and for the combined one. 
The four catalogs obtained were then  combined by matching the Cartesian 
position of the objects, with a tolerance of 0\arcsec.6.

Magnitudes (Sextractor's MAG\_BEST) were measured in the single band images,
and the colors were computed using  4\arcsec~ diameter apertures.
Astrometric and morphological data were obtained from the VRI image.

\begin{figure}[h!]
\plotone{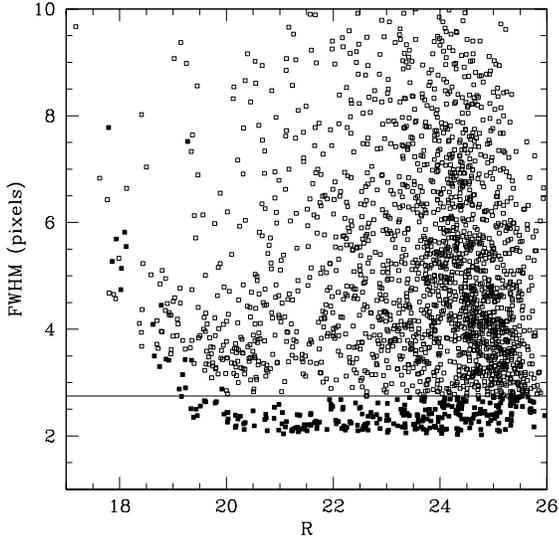}
\caption
{Magnitude - FWHM diagram for objects in the field of A1451.
Filled squares represent objects classified as point-like sources, following the
criteria described in the text. 
The continuous line is the threshold FWHM adopted for star/galaxy separation
of objects fainter than R=20. The star sequence for unsaturated objects is 
clearly identifiable below the line.
\label{fwhm}}
\end{figure}

We have produced three sub-samples of detected objects:
stars, cluster galaxies, and other galaxies.
Stars have been selected using two different criteria. For unsaturated 
objects (R$\simgreat 20.0$ mag) we have used the FWHM of their light 
profiles. In Figure
\ref{fwhm} we show an example of the distribution of this parameter as a
function of R magnitude, which shows that stars can be easily separated from
galaxies down to very faint magnitudes. For brighter objects, we have used the
{\sc SExtractor} neural network classifier.

\begin{figure}[h!]
\plotone{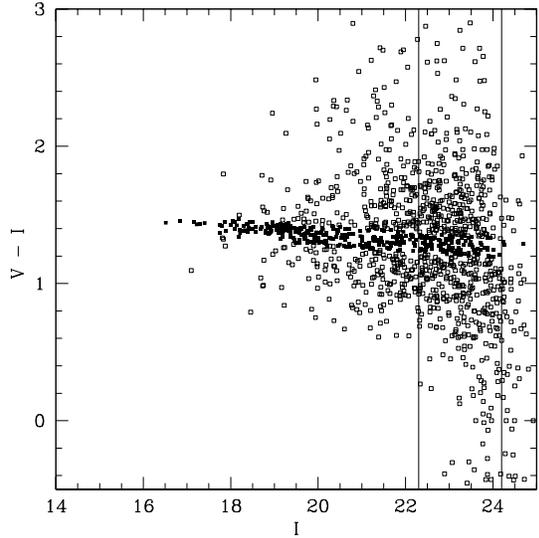}
\caption
{Color-magnitude diagram for galaxies in the field of A1451. Galaxies belonging
to the red cluster sequence are shown as filled squares. The vertical lines 
shows the range of magnitude for the galaxies used for the lensing analysis. 
\label{cormag}}
\end{figure}

In the absence of spectroscopic redshifts for most of the galaxies in the
clusters, the cluster galaxies were selected from the 
color-magnitude diagram $(V-I) \times I$ for each cluster (Figure 
\ref{cormag}). In this type of diagram, the early-type cluster
galaxies occupy a well defined locus, the so called red cluster sequence, with
almost the same color. After determining the typical color of the ellipticals
in each cluster, we selected as cluster galaxies those with colors within a
strip of width 0.2 mag matching the red sequence of each cluster.
This selection procedure naturally will not include the bluer galaxies belonging
to a particular cluster and having colors
outside the color range defined by the strip in the cluster
color-magnitude diagram. These bluer galaxies not recognized
as cluster members cannot be distinguished from the background galaxies
needed for  the weak-lensing analysis and thus may introduce a certain
level of ``noise''. We will return to this issue later.

\subsection{Shape Measurements}

Weak-lensing analysis needs careful measurements of the ellipticity of
background galaxies observed in the field of a cluster of galaxies.

We made galaxy shape measurements, including the correction of the seeing
circularisation and PSF anisotropies, using the {\sc im2shape} software
\citep{im2shape}. 
Following a proposition by \citet{K99}, this software models
astronomical objects with a sum of Gaussians with elliptical base.
Let ${\bf x}$ be a position on a CCD image. The intensity of a 
galaxy as a function of position is modeled as
\eq
I({\bf x}) =  \sum_i {A_i \over 2\pi|C_i|}
e^{-({\bf x}-{\bf x_i})^T C_i({\bf x}-{\bf x_i})/2}
\eeq
where the ${\bf x_i}$, $A_i$, and $C_i$ are, respectively, the center of
each Gaussian, its amplitude, and its covariance matrix. The components
of $C_i$ can be written is terms of the ellipse parameters 
$a$, $b$,  and $\theta$, that are its semi-major and semi-minor axes and the
position angle of the major axis, respectively. The ellipticity, following 
the convention usually adopted in lensing studies, is defined as
\eq
\epsilon = {a-b \over a+b}
\eeq

At this point it is convenient to define the ellipticity projected 
tangentially to the direction of a predefined cluster mass center,
\eq
\epsilon_t = \epsilon \cos{[2(\theta - \phi)]}
\eeq
where $\phi$ is the azimuthal coordinate of the galaxy.  

{\sc im2shape} uses a Bayesian approach to determine these parameters, with
carefully calculated uncertainty estimates. It also deconvolves
the measured shape using a PSF that is also given as a sum of Gaussians,
so that the  deconvolution process can then be done in a fully analytical way.

The steps actually performed in measuring galaxy shapes on our images can be
summarized as follows.
First, {\sc im2shape} is used to model images of the unsaturated stellar 
objects as single Gaussians.
These stars (in fact stellar-like objects) are divided in four sets, following 
their positions regarding the CCD quadrants. 
Stars with $\epsilon$, $\theta$ or base area unrepresentative of their
respective quadrant averages are removed through a 2-sigma clipping
process. 
This cleaned star catalog is used to map the PSF along the field.
After, the same program is applied to the galaxies in the CCD image. 
Each galaxy image is modeled as a sum of two
Gaussians, which is adequate for exponential and de Vaucouleurs profiles,
given typical galaxy sizes, CCD pixel-sizes and PSFs
\citep[see][for a discussion]{K99}. The parameters $x$, $y$, $\epsilon$,
and $\theta$ are forced to be the same for both Gaussians.
Using the five nearest stars, a local PSF is calculated for the
position
of each galaxy and a deconvolution of their images is performed. 
The result of this procedure is a list with the elliptical parameters of the
galaxies in a cluster field. 

\subsection{Weak-Lensing Sample Selection}

A key step in weak-lensing analysis is the selection of galaxies that
will be
used as probes of the gravitational field of the lensing cluster.
These galaxies must
be behind the cluster and we should be able to measure their
shapes with good accuracy. For this, it is necessary to have distance 
estimates for all objects in the field. It is possible to select background
galaxies through photometric redshifts \citep[e.g.][]{Athreya}, but an
efficient application of this technique 
requires more abundant color information than what we have.
In the absence of redshift information, we proceeded by building
a master catalog of putative background galaxies, selected
by flux in the non-cluster galaxy catalog. We selected only 
galaxies with absolute R-band magnitude fainter than 
-16.6  at the cluster distance \citep[roughly M$^*$+5;][]{Goto}. 

The final catalog is obtained after removal of a
number of objects, obeying the following criteria.
Initially, all objects inside a radius  of 1 arcmin
around the cluster 
center were excluded from the lensing sample \citep[following, for
example, ][]{hoekstra2002}.
The reason for this is twofold. First, in
these regions (mainly the inner 30\arcsec) mass densities are closer to the
critical value, and then the weak-lensing regime is no longer a good
approximation. Second, in central regions there is a
high surface density of low luminosity cluster components and diffuse
light, that can contaminate the weak-lensing sample and bias the shape 
measurements.

Relevant cases of light contamination by luminous neighbors were detected
by comparing differences in the object center estimated by {\sc SExtractor} and
by {\sc im2shape}, because the first has a deblending procedure, while the
second does not. 
Thus, objects with differences between center estimates greater
than 3 pixels were also removed from the sample.  
Finally, all galaxies with uncertainties in $\epsilon_t$ larger
than 0.35 were also removed. We found that this value is
a good compromise between the total number of galaxies in the final
sample
and the accuracy of shape measurements. It should be stressed,
however,  that the results  presented here depends only weakly on this value, 
since in the whole analysis
the inverse of the square of the uncertainties are used as statistical
weights (see for example Section \ref{fit}).

The total number of galaxies useful for shear measurements is presented in
column (2) of Table \ref{resultstable}. Their surface density
ranges from 4.7 to 17.8 gal/arcmin$^2$.
This broad range are due to cosmic variance, almost unavoidable differences in
the deepness of the images, (see the values of the seeing and $\mu_{sky}$ in
Table \ref{VMACS}) and also to the different bright apparent magnitude cutoffs
for  the galaxy lensing samples.

\subsection{Mass and Light Maps}

In the weak-lensing regime, the mass distribution can be obtained from
the pattern of distortion of galaxy shapes at a position ${\bf x}$,
${\langle\epsilon_t  ({\bf x}) \rangle}$,
that depends only of
the reduced shear $g= \gamma /(1-\kappa)$, 
where $\gamma$ and $\kappa$
are the shear modulus and convergence at ${\bf x}$.
The latter is proportional to the mass density and is defined as:
\eq
\kappa = {\Sigma ({\bf x}) \over \Sigma_c},
\eeq
where the critical surface density $\Sigma_c$ is given by
\eq 
\Sigma_c = {c^2 \over 4 \pi G} {D_s \over D_l D_{ls}} 
\eeq
and $D_l$, $D_s$, and D$_{ls}$ are the angular diameter distances from
the observer to the lens, from the observer to the source, and from the
lens to the source.

We have reconstructed the two-dimensional mass density distributions
from distortion maps using the {\sc LensEnt} software \citep{sarah1,phil}. 
This software computes $\kappa$ maps using a
maximum-entropy method, taking each background galaxy image shape as an
independent estimator of the
reduced shear field. This map is smoothed by a Gaussian function, called
intrinsic correlation function (ICF), whose
FWHM is determined by Bayesian methods. The program avoids any
binning of the data and allows the reconstruction of complex mass 
distributions.
We present in column (3) of Table \ref{resultstable} the optimal
ICF-FWHM for the smoothing function found by LensEnt.

The resultant mass maps for all clusters,
with $64 \times 64$ pixels across, can be seen in Appendix \ref{figs}.
In order to minimize edge effects, LensEnt produces maps in 
a grid twice as big as the data region.
Here we only show the part of a map corresponding to the field actually
observed. 

It should be noted that since this method takes into account only galaxy
distortions and
ignores their amplifications, the convergence determined this way will suffer
from the ``mass sheet degeneracy'': the distortion pattern can be reproduced
by any mass distributions that obeys the 
transformation $\kappa \rightarrow \lambda\kappa +(1-\lambda)$, where
$\lambda$ is a real number.

To estimate the significance of the features found in mass maps we 
have also performed 100 bootstrap realizations for each cluster, adding 
to galaxy ellipticities Gaussian noise with amplitude equal to their
observational errors. From the mass maps produced in the simulations,
we computed a standard deviation ($\sigma$) map and, by dividing the mass map 
by the $\sigma$-map, we obtained a significance map. 
The significance maps show that most clusters presents individual pixel
signal-to-noise ratios ranging between 1 and 3, with values as high as 8
being found in some cases.

We have also computed light maps (also shown in Appendix  \ref{figs}) of 
the light distribution of galaxies in the red sequence.
The field is divided in $32 \times 32$ pixels and the smoothed light
density of each pixel is estimated as
\eq
{\cal L} \propto \sum_i  L_i~e^{-{d_i^2\over 2\sigma^2}}
\eeq 
where  $L_i$ is the luminosity of the
$i$-th galaxy, and $d_i$ is its angular distance to the pixel center. 
The value of $\sigma$ is chosen 
to have the same FWHM of the ICF of the mass map.

For two clusters, A1651 and A1664, the algorithm was not able to
reconstruct the density distribution. This will be
discussed in Section \ref{fail}.

\begin{figure}[h!]
\plottwo{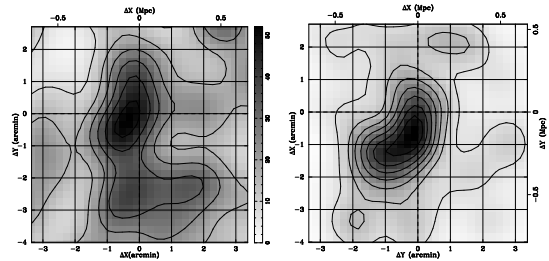}{f3b.eps}
\caption
{Mass (left) and Light (right) maps for the cluster A1451.
The mass map is produced using the distortions of the faint background
galaxies as probes of the mass density of the cluster. The gray level
scale shown at the right edge of the map is in unities of the critical
density. The light map is the 
result of the smoothing of the magnitude of the clusters members 
belonging to the red sequence (see text for details).\label{maps}}
\end{figure}

\subsection{Profile Fitting\label{fit}}

In order to avoid the mass sheet degeneracy and to obtain quantities that 
can be compared with data determined by other methods, 
we fit physically motivated parametric models to the distortion maps. We
considered tho different mass models: a singular isothermal sphere (SIS)and
a singular isothermal ellipsoid (SIE).

Let us assume a polar reference system with origin
at the cluster center, so that any point on the image can be represented by
an angular radial coordinate $\theta$ and an azimuthal angle $\phi$.
For the SIS profile the convergence and the shear are given by
\eq
\kappa= \gamma = {1 \over 2} {\theta_E \over \theta}
\eeq
where $\theta_E$ is the Einstein
radius, which is related to the one-dimensional velocity dispersion of the
isothermal sphere ($\sigma_{SIS}$) as
\eq
\theta_E = 4 \pi {\sigma_{SIS}^2 \over c^2} {D_{ls} \over D_s}.
\label{reSIS}
\eeq
The SIE profile \citep[e.g.,][]{Kormann} has properties similar to those
of SIS models, but the values of the shear and the convergence are now
given by
\eq
\kappa= \gamma = {1 \over 2}  {\theta_E \over \theta} f~
[\cos{(\phi - \alpha)} + f^2 \sin{(\phi - \alpha)}]^{-1/2}
\eeq
where $f<1$ is the axial ratio $b/a$, $\alpha$ is the position angle 
of the ellipsoidal matter distribution and $\phi$ the azimuthal coordinate.
For this model, in the weak-lensing regime ($\kappa<<1$), the
shear is oriented tangentially to the direction to the mass center.

The SIS model has only one free parameter,  $\theta_E$, whereas the SIE
has three free parameters: $\theta_E$, $f$, and $\alpha$. In both cases the
position of the cluster center were defined in advance with the help of the
mass maps. For each model, the best-fit parameters were obtained through
minimization of the $\chi^2$ statistic, defined as:
\eq
\chi^2 = \sum_i{(\epsilon_{t,i} - g_{t,i})^2\over
             {\sigma_{{\epsilon_t},i}^2+\sigma_\epsilon^2}},
\eeq 
where $\epsilon_{t,i}$ and $\sigma_{{\epsilon_t},i}$ are the tangential
ellipticity and its error for the $i$-th galaxy, 
$\sigma_\epsilon=0.3$ \citep{mellier99} is the dispersion associated with
galaxy intrinsic shapes, and $g_{t,i}$ is the tangential reduced shear at
the position of this galaxy, which quantifies the ellipticity induced by the lensing 
distortion. For all models we consider here it may
be assumed that, in the weak lensing regime, $|\vec{g}| = g_t$
This method of profile fitting is similar to the one developed 
by \citet{king&s} and has the advantage of avoiding data binning.

The mass center was defined in general as the center of the mass
map.
When there is consistency between the peak of the mass map and a bright cluster
galaxy (the dominant one in most cases), the latter one was used as the center
position. In some cases (A2104, A3739 and A3888) the maximum of the mass map and
therefore the adopted mass center, is in a position between bright cluster
members, and in other cases (A85, A2811 and A1650) in a position uncorrelated with 
the cluster galaxies.

The best-fit parameters of SIS and SIE models are also shown in Table
\ref{resultstable}. We have estimated $\sigma_{SIS}$ ($\sigma_{SIE}$ for the
elliptical model) from $\theta_E$ using
for each cluster an average value of $D_{ls}/D_s$ obtained in a way
analogous to that adopted by \citet{dahle}. Considering the
minimum and average values of I magnitudes in our weak-lensing sample,
we selected sub-samples  of HDF  galaxies \citep{soto} and, from
their photometric redshifts, the mean value of the distance ratio 
above was calculated. 
In this table we also show the average redshift of the galaxies used in the
lensing analysis with their respective uncertainties (that are not taken into
account on the estimate of the uncertainties in $\sigma_{SIS}$).

When fitting noisy data or clump mass distributions with the SIE profile, one 
often obtain unrealistic low values of the axial ratio ($f<0.3$). 
Clusters where this happened have an empty entry in Table \ref{resultstable}.
We dealt with it adopting a constant value of $f=0.3$ when the 
fit gives values smaller than this one.
On the other hand, the value obtained
for the position angle is quite robust, being rather insensitive to 
$f$. 
Soft core (i.e. non singular) and \citet{NFW} profiles have parameters,
such as the core radius or the NFW concentration parameter, whose determination 
depends critically on information from the inner parts of the cluster.
Since we do not have useful data in the central arcmin, or even at larger radius,
(that are important for NFW profiles), we are unable to fit properly our data using
these models.

Another point that deserves mention in Table \ref{resultstable} are the small
values of $\chi_{red}^2$. This happens due to the large value of
the intrinsic shape dispersion $\sigma_\epsilon$.

\begin{figure}[h!]
\plotone{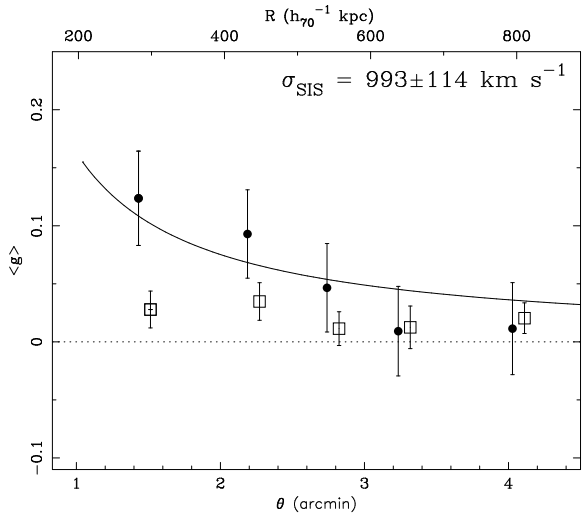}
\caption
{The shear profile for the cluster A1451. The filled symbols
correspond to average ellipticies of the faint background galaxies projected
tangentially to the cluster center. Each point represents nearly 1/5 of the
galaxies. The open squares are the same, but of ellipticies projected in a
direction 45\degr relative to the center. The solid line shows the best fit
SIS profile.
\label{shear_prof}}
\end{figure}

\begin{deluxetable}{lcccccccccc}
\setlength{\tabcolsep}{4pt}
\tabletypesize{\small}
\tablewidth{0pt}
\tablecaption{Best fit parameters \label{resultstable}}
\tablehead{
 &  &  & & \multicolumn{3}{c}{SIS} & \multicolumn{4}{c}{SIE}  \\
\colhead{Cluster} & \colhead{N} &  \colhead{{\tiny ICF-FWHM}} &
\colhead{$\langle z\rangle$} & \colhead{$\theta_E$} &
\colhead{$\sigma_{SIS}$}  &  \colhead{$\chi_{red}^2$}  &
\colhead{$\sigma_{SIE}$} & \colhead{$f$} & \colhead{$\alpha$} &
\colhead{$\chi_{red}^2$}\\
\colhead{(1)} & \colhead{(2)} & \colhead{(3)} & \colhead{(4)} &
\colhead{(5)} & \colhead{(6)} & \colhead{(7)} & \colhead{(8)}&
\colhead{(9)} & \colhead{(10)}& \colhead{(11)}}
\startdata
A2744 & 323& 220   & 1.01$\pm$0.04 & 27.5$\pm$4.3 &$   1491\pm    116$& 1.54  & $   1545\pm    111$ &$0.52\pm 0.21$ &\phs$ \phn1.73\pm 16.65$ & 1.53 \\
  A22 & 554& 240   & 0.84$\pm$0.05 & 10.7$\pm$3.6 &$\phn663\pm    113$& 1.46  & $\phn812\pm \phn79$ &$  	  $ &\phs$ 14.81\pm    10.79$ & 1.43 \\
  A85 & 492& 110   & 0.81$\pm$0.05 & 20.7$\pm$3.8 &$\phn917\pm \phn85$& 1.11  & $\phn926\pm \phn82$ &$0.54\pm 0.27$ &\phs$ 21.37\pm    21.80$ & 1.11 \\
A2811 & 498&  80   & 0.86$\pm$0.05 & 16.3$\pm$5.6 &$\phn863\pm    149$& 1.13  & $\phn861\pm    134$ &$  	  $ &	 $-34.38\pm    15.22$ & 1.13 \\
A1437 & 385& 160   & 0.84$\pm$0.05 & 10.8$\pm$4.5 &$\phn734\pm    152$& 0.95  & $\phn780\pm    127$ &$  	  $ &\phs$ 45.94\pm    16.74$ & 0.94 \\
A1451 & 462&  80   & 0.90$\pm$0.05 & 16.8$\pm$3.9 &$\phn993\pm    114$& 0.95  & $   1087\pm \phn98$ &$            $ &	 $-14.60\pm \phn9.80$ & 0.94 \\
A1553 & 178& 160   & 0.81$\pm$0.05 & 15.4$\pm$6.2 &$\phn923\pm    186$& 1.05  & $\phn925\pm    164$ &$  	  $ &	 $-49.04\pm    19.70$ & 1.05 \\
A1650 & 279& 160   & 0.80$\pm$0.05 & 17.7$\pm$5.4 &$\phn876\pm    133$& 1.08  & $   1033\pm    109$ &$  	  $ &	 $-29.03\pm    10.36$ & 1.06 \\
A1651 & 391&\nodata& 0.73$\pm$0.04 &  1.0$\pm$4.9 &$\phn210\pm    498$& 0.91  & $\phn489\pm    116$ &$  	  $ &\phs$ 70.72\pm    47.99$ & 0.91 \\
A1664 & 246&\nodata& 0.80$\pm$0.05 & -1.9$\pm$6.3 &  \nodata	      &\nodata& $\phn402\pm    264$ &$  	  $ &	 $-74.14\pm    92.74$ & 1.05 \\
A2029 & 549& 140   & 0.83$\pm$0.05 & 25.2$\pm$3.1 &$   1039\pm \phn64$& 0.86  & $   1052\pm \phn63$ &$0.62\pm 0.21$ &	 $-26.11\pm    19.63$ & 0.86 \\
A2104 & 378&  90   & 0.86$\pm$0.05 & 20.7$\pm$4.3 &$   1039\pm    108$& 0.86  & $   1038\pm    102$ &$0.39\pm 0.22$ &\phs$ 58.44\pm    14.44$ & 0.86 \\
A2163 & 261& 240   & 0.89$\pm$0.04 & 17.6$\pm$5.0 &$   1021\pm    146$& 0.90  & $   1094\pm    125$ &$  	  $ &	 $-83.53\pm    12.50$ & 0.88 \\
A2204 & 347& 110   & 0.86$\pm$0.05 & 20.3$\pm$4.1 &$   1028\pm    104$& 0.81  & $   1035\pm    102$ &$0.58\pm 0.32$ &\phs$ \phn9.73\pm 27.80$ & 0.81 \\
A3695 & 294& 130   & 0.83$\pm$0.05 & 19.5$\pm$4.9 &$\phn928\pm    117$& 1.10  & $\phn987\pm    104$ &$            $ &	 $-23.08\pm    11.08$ & 1.09 \\
A3739 & 229& 110   & 0.89$\pm$0.05 & 26.2$\pm$4.9 &$   1203\pm    112$& 1.06  & $   1210\pm \phn96$ &$  	  $ &\phs$ 31.71\pm \phn7.37$ & 1.05 \\
A2345 & 364& 150   & 0.89$\pm$0.05 & 15.0$\pm$4.6 &$\phn909\pm    138$& 1.08  & $\phn965\pm    126$ &$            $ &\phs$ 87.18\pm    14.14$ & 1.08 \\
A2384 & 420& 150   & 0.80$\pm$0.05 & 12.2$\pm$4.2 &$\phn737\pm    126$& 1.05  & $\phn787\pm    108$ &$  	  $ &\phs$ 18.97\pm    14.18$ & 1.05 \\
A2426 & 343& 210   & 0.80$\pm$0.05 & 10.1$\pm$4.8 &$\phn676\pm    158$& 0.91  & $\phn767\pm    132$ &$  	  $ &\phs$ 58.74\pm    18.14$ & 0.90 \\
A3856 & 429& 110   & 0.87$\pm$0.05 & 20.1$\pm$4.0 &$   1005\pm \phn97$& 1.04  & $   1022\pm \phn93$ &$  	  $ &	 $-33.40\pm \phn8.86$ & 1.03 \\
A3888 & 270& 130   & 0.86$\pm$0.05 & 18.6$\pm$4.7 &$\phn981\pm    125$& 1.21  & $   1008\pm    125$ &$0.58\pm 0.40$ &\phs$ 79.13\pm    34.82$ & 1.22 \\
A3984 & 607& 210   & 0.95$\pm$0.05 & 22.2$\pm$3.1 &$   1093\pm \phn77$& 1.27  & $   1150\pm \phn66$ &$            $ &\phs$ 55.19\pm \phn6.38$ & 1.25 \\
A2597 & 524& 150   & 0.89$\pm$0.04 & 14.1$\pm$3.7 &$\phn776\pm    101$& 1.03  & $\phn853\pm \phn79$ &$  	  $ &	 $-18.57\pm    10.59$ & 1.01 \\
A4010 & 673& 160   & 0.87$\pm$0.04 & 11.3$\pm$3.3 &$\phn706\pm    104$& 1.27  & $\phn776\pm \phn84$ &$  	  $ &\phs$ 56.98\pm    11.95$ & 1.26 \\
\enddata
\tablecomments{(1) Cluster name. (2) Number of galaxies used for weak-lensing
analysis. (3) FWHM of the smoothing function used to build the 
mass map in arcsec. 
(4) Average redshift and standard deviation of the galaxies in
the weak lensing sample (see text for details).  
(5,6) The Einstein radius (in arcsec) and $\sigma_{SIS}$ and respective
uncertainty obtained by the fitting of a SIS profile. (7) The $\chi^2$ of the
SIS fit
divided by the number of degrees of freedom. (8),(9),(10) $\sigma_{SIE}$, axial
ratio and position angle (counted counterclockwise with the North direction
been the
origin) and respective uncertainties obtained by the fitting of a SIE profile.
When the best SIE fit gives a value of f smaller than 0.3,
we have done the fit again using a fixed value of f=0.3 and let the entry for f
in the table empty.
(11) The $\chi^2$ of the SIE fit divided by the number of degrees of freedom.}
\end{deluxetable}



\section{Discussion}

\subsection{Mass and Light Distributions}

Although each galaxy cluster of our sample has particular features, as can be seen
in the display of the mass and light distributions provided in the Appendix,
there are some general trends and correlations that can be drawn from
these maps. A detailed and more quantitative comparison of these
distributions will be addressed in a future paper.
 
In most cases (17 out of 22), the center of mass of the cluster
 (nearly) coincides with the
position  of the brightest cluster galaxy (BCG)
or the main galaxy clump. We are including in this category the clusters
A22, A1437, and A2163, where the mass maps show peaks near the
position of the central galaxy, despite the presence of more prominent
peaks near the edges of the frame (see relevant figures in the Appendix).

Although most of the clusters mass maps show single clumps, an effect that
can be attributed in part to poor resolution, some of them present
secondary peaks (e.g. A1451, A3695), or extensions of the main peak in the
direction of substructures
present in the light distribution (A2744, A2104, A2597 and A3739).
Conversely, the  clusters
A1553 and A2163 present prominent light clumps without counterparts in the
mass map.

For some cluster, the mass and light maps 
suggest the existence of ``dark clumps'' (mass concentrations without
optical counterparts), but their reality is doubtful since most of them 
are near the edge of the observed cluster field. Nevertheless, in the map of A1451 
(see Fig. \ref{a1451fig}) there is a significant structure, South
of the main cluster clump, with an extension to the West, that is not too close
to the frame border and is not associated with bright cluster galaxies.
Indeed, our bootstrapping of galaxy ellipticities shows that this is a robust
feature. 
The X-ray isophotes of \citet{valtchanov} also do not show any emission 
excess at this region.
The nature of this feature, whether it is a real dark clump belonging
to A1451 or a background massive structure, will be subject of future
investigations.

Dark matter profiles tend to present significant departures from circular
symmetry. This can be noticed by the smaller values of $\chi^2_{red}$ 
presented by SIE in relation to the SIS fits, as well as by the uncertainties of
$\sigma_{SIE}$, that are $\sim$20\% smaller for the non-circular profile. 
Furthermore, the mass maps often show elongated distributions, with major
axis with position angles similar to those given by the SIE fitting.
Sometimes the ellipticity inferred from the mass map is smaller than that from
the SIE fitted profile. This is due to two reasons: first,
the smoothing process in the mass map calculation tends to circularize 
features whose scales are smaller than the map resolution;
second, as discussed before there is a bias toward small ellipticity
values in the fitting of SIE profiles with noisy data. 

\begin{figure}[h!]
\plotone{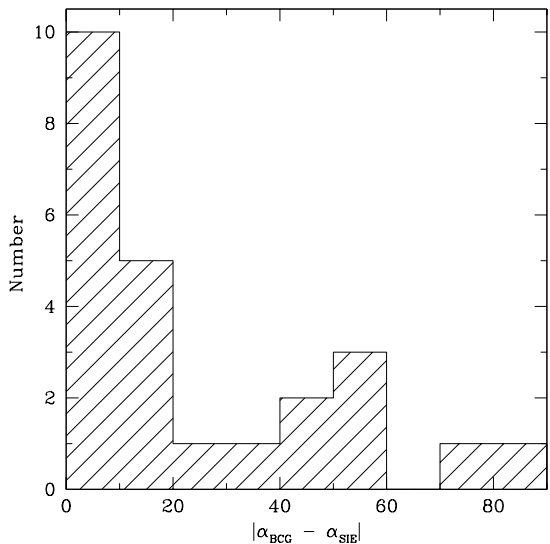}
\caption{Alignment between the position angle of
the brightest cluster galaxy $\alpha_{BCG}$ and of the mass
distribution
$\alpha_{SIE}$. For clusters without a clear dominant galaxy, the direction
linking the two first ranked cluster members is used instead.\label{pa}}
\end{figure}

A very robust result that comes from the SIE fits is that the major-axis of 
the mass distribution  is clearly aligned with the major-axis of BCGs. This
is shown in Fig. \ref{pa}, where we present the absolute value of the
difference between the position angle of a BCG (from {\sc SExtractor})) and 
the position angle given by the SIE mass model. For clusters without a clear 
dominant galaxy (A2744, A1451, A3888, A3984), the direction linking 
the two first ranked cluster members 
was adopted to characterize the orientation of the light distribution.
On the other hand, for at least two
cases (A1451, A2426) the main direction of the dark matter distribution is more
related with the direction linking bright cluster members than with the
direction of the BCG. 

The strong alignment between light and mass in Mpc scales found
here has also been detected, at smaller scales, by the excess of strong
lensing features found in the direction of the major axis of BCGs 
\citep{luppino, paper2}.

Considering that alignments between several cluster components (BCGs - cluster
galaxies - sub-structures -  surrounding large-scale structures) have already
been found \citep[e.g.][]{west,plionis}, the present result reinforces the
conclusion that clusters, in the same way as their  dominant galaxies, 
have been formed by the anisotropic merger of smaller structures along
large-scale filaments, in close 
agreement with numerous cosmological simulations based on the CDM
paradigm \citep[e.g.][]{virgo}. 
It should be stressed here that the diffuse luminous halo
associated with a cD galaxy tends to produce the opposite effect. 
This halo will produce a significant gradient in the background of a
certain galaxy, that contaminates the measure of its shape, causing a 
stretch of it in the radial direction.
As this halo is brighter in the direction of the major-axis, 
galaxies located in
this direction will be more affected.
This is probably what is happening in A2029, 
which has a very luminous diffuse component \citep{uson2029} and where
the major-axis of the mass distribution is about 60 degrees of the
central cD major-axis.
The hypothesis of light contamination in this case is supported by the fact that
when the central exclusion radius is increased from 60\arcsec~ to
75\arcsec, it produces a major change in $\alpha$, that becomes almost parallel
to the BGC main axis, however without causing 
appreciable changes in $\sigma_{SIE}$. We did not detect such effect in any
other cluster of the sample. 

\subsection{The dynamical state of massive clusters} 

The dynamical state of the clusters in the present sample can be accessed
through a comparison between the mass estimates made by weak-lensing, 
dynamical, and X-ray techniques. The key idea is that the weak-lensing
signal is independent of the cluster dynamical state, whereas 
other techniques rely on assumptions of equilibrium.

\begin{deluxetable}{lcccccrc}
\tablewidth{0pt}
\tablecaption{
\label{literatura}}
\tablehead{
\colhead{Nome} & \colhead{$k$T$_{SIS}$} & \colhead{$k$T$_{SIE}$} & 
\colhead{$k$\tx } & \colhead{Refs.} & \colhead{$\sigma_v$} & \colhead{N}& \colhead{Refs.} \\
\colhead{} & \colhead{(keV)} & \colhead{(keV)} & 
\colhead{(keV)} & \colhead{} & \colhead{ (\kms)} & \colhead{} \\
\colhead{(1)} & \colhead{(2)}& \colhead{(3)}& \colhead{(4)} &
\colhead{(5)}& \colhead{(6)}& \colhead{(7)}& \colhead{(8)}}
\startdata
A2744 & 14.13$^{+2.28}_{-2.11}$ & 15.17$^{+2.26}_{-2.10}$& 11.04$^{+0.49}_{-0.45}$ &(a)  & 1777$^{+151}_{-125}$ &  55   & (g) \\
A22   &  2.79$^{+1.03}_{-0.87}$ &  4.26$^{+0.78}_{-0.73}$& \nodata		   &     &  693$\pm$251	        &   7   & (h) \\
A85   &  5.34$^{+1.04}_{-0.94}$ &  5.45$^{+1.00}_{-0.92}$&  6.51$^{+0.16}_{-0.23}$ &(b)  & 1097$^{+76}_{-63}$	& 305   & (i) \\
A2811 &  4.73$^{+1.78}_{-1.49}$ &  4.76$^{+1.55}_{-1.33}$&  5.31$^{+0.17}_{-0.16}$ &(c)  &  695$^{+200}_{-108}$ &  13   & (j) \\
A1437 &  3.42$^{+1.56}_{-1.27}$ &  3.87$^{+1.30}_{-1.11}$& \nodata		   &     & \nodata 	        &\nodata&     \\
A1451 &  6.27$^{+1.52}_{-1.36}$ &  7.47$^{+1.40}_{-1.28}$& 13.40$^{+1.9}_{-1.5}$   &(d)  & 1338$^{+130}_{-90}$  &  37   & (d) \\
A1553 &  5.42$^{+2.40}_{-1.96}$ &  5.37$^{+2.00}_{-1.69}$&  9.16$^{+1.02}_{-0.64}$ &(c)  & \nodata	        &\nodata&     \\
A1650 &  4.88$^{+1.60}_{-1.37}$ &  7.25$^{+1.54}_{-1.39}$&  5.68$^{+0.30}_{-0.27}$ &(b)  & \nodata	        &\nodata&     \\
A1651 &  0.28$^{+2.90}_{-0.25}$ &  1.85$^{+1.61}_{-1.11}$&  6.22$^{+0.45}_{-0.41}$ &(b)  & 695$^{+200}_{-108}$  &  62   & (k) \\
A1664 &  \nodata                &  1.58$^{+1.29}_{-2.32}$& \nodata		   &     & \nodata 	        &\nodata&     \\
A2029 &  6.86$^{+0.87}_{-0.83}$ &  7.03$^{+0.87}_{-0.82}$&  7.93$^{+0.39}_{-0.36}$ &(b)  & 1164$^{+98}_{-78}$	&  93   & (k) \\
A2104 &  6.86$^{+1.50}_{-1.35}$ &  6.85$^{+1.41}_{-1.28}$&  9.13$^{+0.69}_{-0.45}$ &(c)  & 1201$\pm$200         &  51   & (l) \\
A2163 &  6.63$^{+2.03}_{-1.76}$ &  7.55$^{+1.78}_{-1.60}$& 12.3 $^{+1.3}_{-1.1}$   &(e)  & 1698 	        &       & (m) \\
A2204 &  6.72$^{+1.49}_{-1.29}$ &  6.81$^{+1.41}_{-1.28}$&  6.38$\pm$0.23	   &(b)  & \nodata 	        &\nodata&     \\
A3695 &  5.50$^{+1.47}_{-1.30}$ &  6.17$^{+1.38}_{-1.24}$&  6.67$^{+2.84}_{-1.99}$ &(b)  &  779$^{+67}_{-49}$	&  96   & (k) \\
A3739 &  9.20$^{+1.79}_{-1.63}$ &  9.41$^{+1.51}_{-1.40}$& \nodata		   &     & \nodata 	        &\nodata&     \\
A2345 &  5.25$^{+1.72}_{-1.47}$ &  5.93$^{+1.62}_{-1.42}$& \nodata		   &     & \nodata 	        &\nodata&     \\
A2384 &  3.45$^{+1.28}_{-1.08}$ &  4.02$^{+1.11}_{-0.97}$& \nodata		   &     & \nodata 	        &\nodata&     \\
A2426 &  2.90$^{+1.52}_{-1.20}$ &  3.85$^{+1.36}_{-1.15}$& \nodata		   &     &  846$\pm$100	        &  15   & (n) \\
A3856 &  6.37$^{+1.34}_{-1.21}$ &  6.74$^{+1.24}_{-1.14}$& \nodata		   &     &  729$\pm$142	        &  22   & (h) \\
A3888 &  6.12$^{+1.66}_{-1.46}$ &  6.46$^{+1.70}_{-1.50}$&  8.46$^{+3.6}_{-2.53}$  &(b)  & 1102$^{+137}_{-107}$ &  50   & (g) \\
A3984 &  7.59$^{+1.11}_{-1.03}$ &  8.30$^{+1.02}_{-0.96}$& \nodata		   &     & \nodata 	        &\nodata&     \\
A2597 &  3.83$^{+1.06}_{-0.93}$ &  4.56$^{+0.82}_{-0.75}$&  4.40$^{+0.4}_{-0.7}$   &(f)  & \nodata	        &\nodata&     \\
A4010 &  3.17$^{+1.00}_{-0.87}$ &  3.88$^{+0.81}_{-0.74}$& \nodata		   &     &  625$^{+127}_{-95}$  &  36   & (k) \\
\enddata
\tablecomments{(1) Cluster name. (2),(3) X-ray temperatures inferred from the SIS/SIE
modeling of the shear data. (4) X-ray temperature with 66\% uncertainties
from the literature. (5) References for \tx.  (6) Velocity dispersions found in
the literature. (7) Number of galaxies used to determine $\sigma_v$. 
(8) References for $\sigma_v$.}
\tablerefs{(a) \citet{allen2000} (b) \citet{ikebe} (c) \citet{white2000}
(d) \citet{valtchanov} (e) \citet{Markevitch-Vikhlinin} (f) \citet{Markevitch}
(g) \citet{girardi} (h) \citet{vicky} (i) \citet{durret} (j) \citet{collins}
(k) \citet{fadda} (l) \citet{liang} (m) \citet{squiresa2163}  
(n) \citet{mazure}}
\end{deluxetable}

\begin{figure}[h!]
\plotone{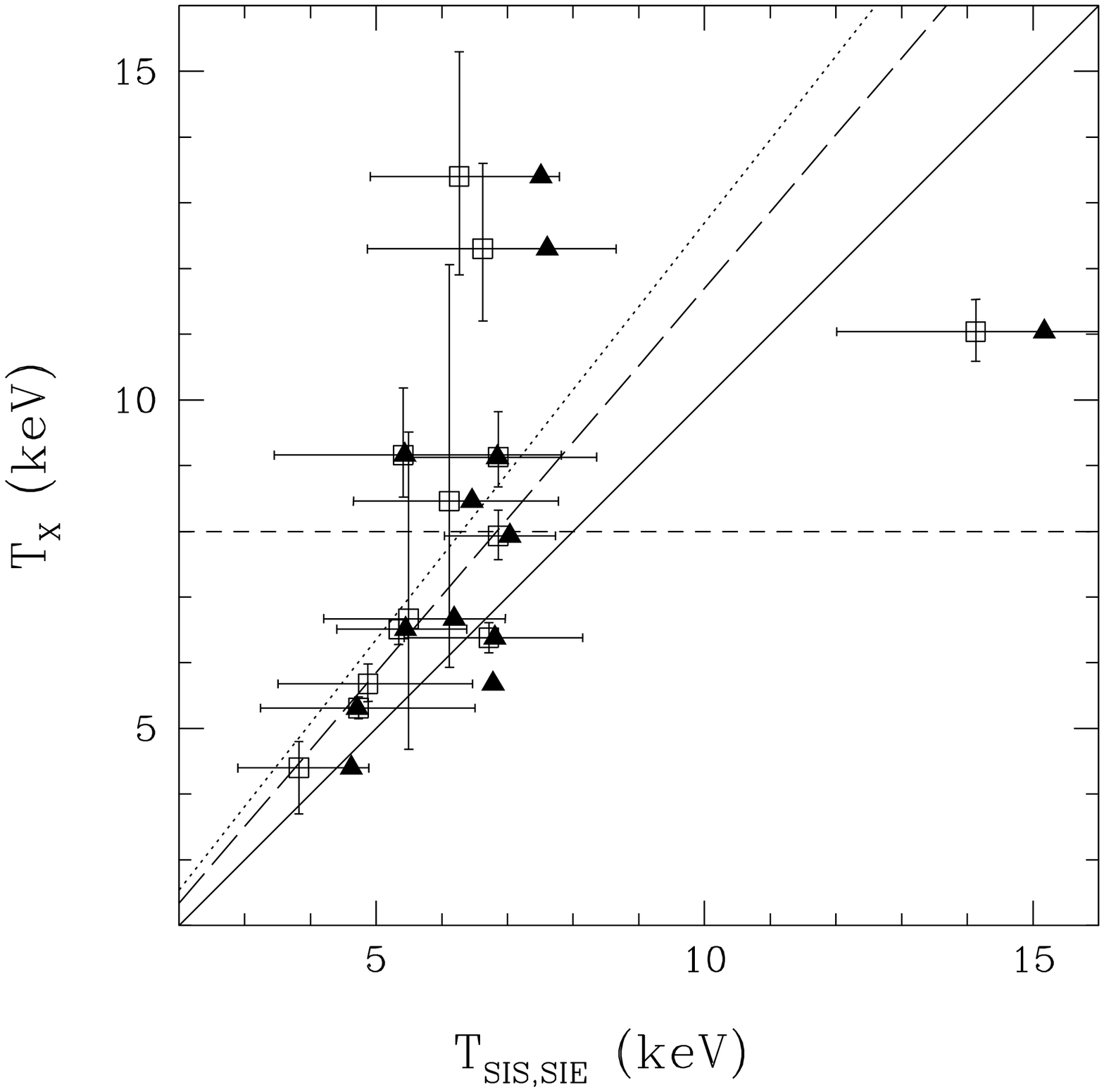}
\caption{Comparison between the ICM temperatures inferred  by the fitting of
isothermal profiles to the shear data, T$_{SIS}$ and T$_{SIE}$, and from X-ray 
measurements, \tx. The squares correspond to the spherical model and
triangles to the elliptical. The error bars of the latter were suppressed 
for clarity. The solid line is defined by T$_{SIS,SIE}$ = \tx.
The dotted and long dashed lines show the best  obtained 
by the SIS and SIE models, respectively, when the origin is kept constant.
The short dashed line indicate T$_X$ = 8 keV. Clusters with higher temperatures 
show signals of dynamic activity. 
\label{temp}}
\end{figure}

\begin{figure}[h!]
\plotone{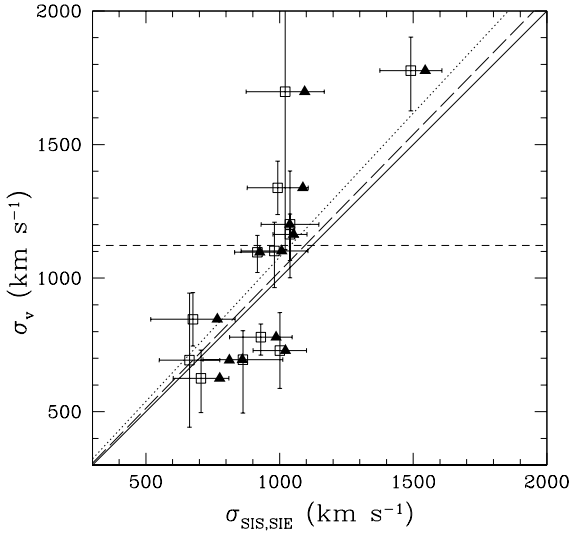}
\caption{Comparison between the velocity dispersion found by fitting 
isothermal profiles to the shear data, $\sigma_{SIE,SIE}$, and those estimated 
through spectroscopic
measurements, $\sigma_v$. The squares correspond to the spherical model and
triangles to the elliptical. The error bars of the latter were suppressed 
for clarity.
The solid line is defined by $\sigma_{SIS,SIE}$ = $\sigma_v$ are
equal. The dotted and long dashed lines show the best fit 
obtained by the SIS and SIE models, respectively, when the origin is kept
constant. The short dashed line indicate $\sigma_v$ = 1122 \kms, which
correspond the gas temperatures of T$_X$ = 8 keV. Clusters with higher
velocity dispersions show signals of dynamic activity.
\label{sig}}
\end{figure}

We have collected measurements of $T_X$ and $\sigma_v$ from the literature
(see  Table \ref{literatura}). Assuming energy equipartition between cluster 
galaxies and gas, we may obtain an estimate of the temperature from our weak-lensing 
determination of $\sigma_{SIS}$ (and for $\sigma_{SIE}$), through the relation \citep[e.g.][]{Sarazin}
\eq
\sigma_{SIS}^2 = {k T_{SIS} \over \mu m_H},
\eeq
where  $\mu=0.61$ is the mean molecular weight and $m_{H}$ the hydrogen mass.
The results are also in Table \ref{literatura}.
In Fig. \ref{temp} we present a comparison between temperatures computed in 
this way with those actually obtained with X-ray observations. In Fig.
\ref{sig} we compare the velocity dispersions determined through lensing
with those obtained directly from galaxy velocities.
The first point to note in these figures is that for many clusters 
the weak-lensing results agree with X-ray or dynamical data 
within 1.5 $\sigma$.  The
clusters A1451, A1553, and A2163, however, show temperatures and velocity
dispersions significantly in excess when compared with estimates derived
from weak-lensing data, suggesting that they should be dynamically active. 
Actually, the analysis of the A2163 temperature map made by
\citet{Markevitch-Vikhlinin} with Chandra data, shows at least two shocked
regions and other evidences that the central region of this cluster is in a
state of  violent motion.  In the same way, \citet{valtchanov}
describes A1451 as being in the final stage of establishing equilibrium after a
merger event, whereas its high X-ray temperature (13.4 keV) would be
 probably due to a shock occurred recently.
A2744 seems to be  an exception, since its temperature is 
significantly lower than
the weak-lensing estimate. The nature of this cluster will be discussed 
in more detail in Section \ref{a2744}.

\begin{deluxetable}{cccc}
\tablewidth{0pt}
\tablecaption{Weighted means and errors of the ratio between cluster temperatures
(X-ray and dynamical) and weak-lensing inferred values.\label{razoes}}
\tablehead{
\colhead{} & \colhead{All} & \colhead{$T_X<8.0$ keV} &  \colhead{$T_X>8.0$ keV}\\
\colhead{} & \colhead{} & \colhead{$\sigma_v<1122$ km s$^{-1}$} &
\colhead{$\sigma_v>1122$ km s$^{-1}$}\\
\colhead{(1)} & \colhead{(2)}& \colhead{(3)}& \colhead{(4)}}
\startdata
$T_X / T_{SIS}$            & $1.27\pm0.09$ & $1.14\pm0.03$ & $1.55\pm0.19$\\
$T_X / T_{SIE}$            & $1.17\pm0.08$ & $1.04\pm0.05$ & $1.42\pm0.15$\\
&&&\\
$(\sigma_v / \sigma_{SIS})^2$  & $1.20\pm0.13$ & $1.01\pm0.14$ & $1.54\pm0.20$\\
$(\sigma_v / \sigma_{SIE})^2$  & $1.11\pm0.12$ & $0.90\pm0.13$ & $1.41\pm0.15$\\
\enddata
\end{deluxetable} 

We present in Table \ref{razoes} the weighted (by the error of the ratio) mean
ratios $T_X/T_{SIS,SIE}$ and $(\sigma_v/\sigma_{SIS,SIE})^2$
for both SIS and SIE mass models, as well as their errors of the mean.
These quantities are proportional to the ratio between the mass computed from
the X-ray emission or cluster dynamics and the mass inferred from weak-lensing.
For the clusters in our sample for which this analysis is possible, we verify  
that the temperature inferred from the lensing analysis is  17\% to
27\% lower than that determined by X-ray observations; the squared velocity
dispersions determined from galaxy velocities are also larger than that
obtained from lensing, by a factor between 8\% and 20\%. These factors are
smaller for the elliptical models.
Other comparisons between dynamical
and weak-lensing mass estimates have been made by \citet{irgens} and 
\citet{smail97}. Whereas the latter work has found values of $\sigma_{SIS}$ 
$\sim$50\% greater than $\sigma_{v}$, the former has found a broad 
consistency between these two quantities.

Dividing the sample in clusters colder and hotter than \tx = 8.0 keV (or,
equivalently, $\sigma_v$ larger or smaller than 1122 \kms), 
the first set presents ratios consistent with unity within the errors, 
whereas the second set shows excess (relative to lensing results) 
of 40-50\%; see Table \ref{razoes}. Not surprisingly, all clusters that we 
have already identified as dynamically active belong to this second set.

These results find a natural explanation in the hierarchical scenario, where
the most massive structures are being formed at this very moment. 
If this is the case, higher ICM temperatures are expected (compared to an
equilibrium state), because the gas is heated by shocks produced by galaxy
groups falling in to the cluster as well as mergers with other clusters
\citep[e.g.][]{Markevitch2002}. 

Higher velocity dispersions are also expected, since the cluster dynamics
will be affected by motion streams and substructures, that broaden the
velocity distribution.

It is worth investigating further whether \tx $\sim$ 8 keV may indeed be
considered a threshold to discriminate between relaxed and active
clusters, because many studies
of the cluster mass function are based on X-ray data, that usually
select the most massive clusters. For example, 11 out of the 106 cluster 
of the sample selected by \citet{xmass} have temperatures equal to or higher 
than 8 keV.
It is also interesting to note that if we use an empirical $\sigma-T$ relation
for clusters as the one of \citet{xueewu}, we find a slightly
better agreement between lensing and X-ray properties; the main trends 
discussed before, however, remain the same.

\subsection{Comparison of $\sigma_{SIS}$ with previous estimates}

Some of the clusters studied here have already  a weak-lensing 
analysis made by other authors. Table \ref{dadosdosoutros} 
compares our $\sigma_{SIS}$ with other works that used the same model.

\begin{deluxetable}{lclc}
\setlength{\tabcolsep}{4pt}
\tablewidth{0pt}
\tablecaption{Comparison of $\sigma_{SIS}$  with other works 
\label{dadosdosoutros}}
\tablehead{\colhead{(1)} & \colhead{(2)}& \colhead{(3)} &  \colhead{(4)}\\
\colhead{Cluster} & \colhead{$\sigma_{SIS}$ (\kms)} &\colhead{$\sigma_{SIS}$
(\kms)} &\colhead{Refs.}\\
\colhead{} &
\colhead{\small Present work} &\colhead{\small Others work}&\colhead{}}
\startdata
A2744  &   $1491\pm116$ &\phn\phn870$^{+41}_{-95}$   & (a)\\
A2104  &   $1039\pm108$ &\phn1390 $\pm$ 180          & (b)\\
A2163  &   $1021\pm146$ &\phn\phn740		  & (c)\\
A2204  &   $1028\pm104$ &\phn\phn950$^{+210}_{-250}$ & (b)\\
A2204  &   $1028\pm104$ &\phn1035$^{+65}_{-71}$      & (d)\\
A2345  &$\phn909\pm138$ &\phn\phn870$^{+260}_{-320}$ & (b)\\
\enddata
\tablecomments{(1) Cluster name. (2) Value of $\sigma_{SIS}$ from 
the literature. 
(3) difference between the value of $\sigma_{SIS}$ determined in this work 
and the value found in the
literature. (4) Reference of the other weak-lensing measurement.
}
\tablerefs{(a) \citet{smail97} (b) \citet{dahle} (c) \citet{squiresa2163}
(d) \citet{clowea2204}}
\end{deluxetable}  

\subsubsection{A2744}

The largest difference between the results in Table \ref{dadosdosoutros} 
is for A2744. It has a
shear profile very flat (see Figure \ref{a2744fig}).
\citet{smail97} found for this cluster $g \sim 0.2$ inside a radius of 
1$\arcmin$, whereas our data give $g \sim 0.14$ in
$1.0\arcmin<R<3.2\arcmin$. Taking into account that \citet{smail97} have used
in their analysis galaxies fainter than those considered here, 
thus probably more distant, and also closer to the
cluster center, it is not unexpected the detection of higher 
gravitational distortions by them.
Indeed, as discussed in Section \ref{a2744}, this cluster
probably correspond to two structures close to the line-of-sight.
Isothermal laws are inadequate, and the fact that the $\chi^2_{red}$ found 
for this cluster is the highest in this sample confirms it.

\subsubsection{A2104, A2204 and A2345}

There are three clusters studied by \citet{dahle} in common with the present
sample: A2104, A2204 and A2345. 
For the last two, the values of $\sigma_{SIS}$ obtained by these authors are
consistent with the values presented here, but for A2104 there is a large
difference.
A point that is worth mentioning is that these authors have used
the approximation $g \sim \gamma$ instead of 
$g = \gamma / (1-\kappa)$. Adopting the same approximation to fit our data
with the SIS profile, we obtain the following values: 1092, 1086 and 957 \kms,
respectively, instead of 1039, 1028 and 909 \kms. It demonstrates that this
approximation causes an overestimation of the inferred velocity
dispersion by a factor of $\sim$5\%, depending on
the cluster density and of the field size, but this is not the cause of this
difference. 
In fact, \citet{dahle} report problems with their images of A2104. 
For their weak-lensing analysis they used only one I band image, obtained with
1 hour integration with a seeing (FWHM) of 0.7\arcsec~ at the 
University of Hawaii 2.24 m telescope.
Our data for A2104, on the other hand, has been collected in one of the
best nights at VLT in that semester, when the seeing was 0.41\arcsec.
This makes us very confident of our results. 

Our A2104 mass map is also very different from the one obtained by these 
authors. 
Accordingly to them, the mass distribution is nearly
perpendicular to the major-axis of the central D galaxy, which is also along 
the direction of the two brightest galaxies, and is the direction where
the gravitational arcs are seen. However, our mass map as well as
the fitting of SIE models show that A2104 has an elliptical mass distribution 
aligned with the central galaxy major-axis, as expected if the BCG light is a
good tracer of the cluster potential \citep[see for example][]{MFK93}.

\subsubsection{A2163}

Our estimate of the velocity dispersion of A2163 is also much higher than that
reported by \citet{squiresa2163}. These authors have 
observed this cluster in V and I with total expositions of one hour at the
3.6 CFH telescope, under seeing of $\sim$0.8\arcsec. 
They have used $\sim$ 700 galaxies fainter than I=20.5 mag or V=22.0 mag 
for the lensing analysis, in a field of $7 \times 7$ arcmin.

Here, we have used a much smaller sample of background galaxies, in a field 
of comparable size, because we have considered only galaxies fainter than 23.3
R mag, that is roughly 2.0 mag fainter than the bright end limit used 
by Squires and collaborators.
The images of this cluster were taken in  
the same night as those of A2104, when the seeing was exceptional.
A2163 is at a moderately high redshift ($z=0.2$) and lies in a region of the
sky not too far from the galactic plane (b=30.5$\degr$), where the star field
is dense and the galactic absorption is not negligible
\citep[A$_V$ =  1.74;][]{schlegel}, so the weak-lensing sample of 
\citet{squiresa2163} may be more affected by contamination by stars and
foreground and cluster galaxies than the present one. This might be the
cause for a possible dilution of the lensing signal measured by them. 
Indeed, adopting a comparable magnitude cutoff, $R>21.5$ mag, we obtain 
$\sigma_{SIS}=$887$\pm$145 \kms, that is much closer to the
\citet{squiresa2163} result.

\subsubsection{A2204}

Table \ref{dadosdosoutros} shows that our results for A2204
are in very close agreement with those from \citet{clowea2204}. This work 
is based on images taken with the
wide-field camera of the ESO/MPI 2.2m telescope (34\arcmin $\times$ 34\arcmin),
indicating that when the data are well described by an isothermal mass profile,
as is the case of A2204, the small field of the FORS1 camera does not 
introduce relevant systematic errors.

\subsection{The peculiar case of A2744\label{a2744}}
This cluster, also known as AC118, is interesting for several reasons. It is 
the most distant ($z=0.308$) and apparently the most massive cluster in our
sample. 
It is also the only cluster where we have found  gravitational
radial arcs \citep[see][]{paper2},  and has 
a flat shear profile which is very poorly fitted by isothermal models.
Furthermore, its weak-lensing mass is smaller than the virial mass but, 
surprisingly, significantly larger than the X-ray estimate.
We should have in mind that the use of an isothermal profile to fit a shallow
mass distribution tends to underestimate the cluster mass, since the shear
depends more on the density gradient than on the density actual value.
But an important clue for the interpretation of this cluster comes
from the dynamical analysis of \citet{girardi}, who found a velocity
distribution with two superposed peaks, produced by two structures along the
line of sight, with velocity dispersions of 1121 and 682 \kms. 
Weak-lensing, in this case, is sensitive to the total mass, but
single isothermal mass models are clearly inadequate for this system. 
On the other side, the mass inferred from the X-ray analysis will be 
dominated by
the luminous and hotter structure; for $\sigma_v = 1121$ \kms
a value of \tx = 8.0 keV is expected, somewhat smaller than the X-ray 
measurement(see Table \ref{literatura}). In this scenario
the X-ray luminosity is the sum of the luminosities of both structures.
The value of \lx for A2744 is actually very high 
($22.05 \times 10^{44}$ \ergs), ranked in second place in this sample. 
Of course, both \tx and \lx may be increased with respect to their
equilibrium values by shocks between these two substructures and, therefore,
all these results are consistent with the
hypothesis that A2744 is actually two close structures along the line of 
sight or even a pair of clusters in process of merging.

\subsection{Negative weak-lensing detections: A1651 and A1664 \label{fail}}

For only 2 out of 24 clusters (8.3\%) we were unable to reconstruct the
two-dimensional mass distribution. 
This number is comparable to that obtained by \citet{dahle}, who
failed to detect weak-lensing in 1 cluster out of 39 (2.5\%). 
These authors have adopted a sample threshold of \lx = 10$^{45}$ \ergs, 
two times larger than ours. Our two failure cases are clusters with
luminosities below this threshold.

For A1664 we were unable to reconstruct
the mass density distribution and to fit a SIS profile, and we have got a 
very marginal detection when fitting a SIE profile. 
This cluster is at $z=0.13$ and  has a X-ray luminosity
very close to our threshold. Neither the X-ray
temperature nor the velocity dispersion is available in literature for this
object. Moreover only 246 galaxies were used for the weak
lensing analysis (the third smaller number of the sample). It seems
that we were unable to detect the weak-shear because its signal
is small and is dominated by the noise in its distortion map.

The case of A1651 is less clear, since this cluster seems to be more massive
than A1664 (see Table \ref{literatura}). In fact, 
the SIS fit resulted in a very small value for $\sigma_{SIS}$, but the SIE 
fit led to a significant detection, resulting in a value of 
$\sigma_{SIS}=539\pm198$ \kms, consistent with the dynamical velocity
dispersion found in literature, 695$^{+200}_{-108}$\kms \citep{fadda}. 
The position angle of the SIE model is almost coincident with that of the 
BCG major-axis, as found for most of the clusters in our sample.
However, the lensing signal is still too low to allow a successful
mass reconstruction by {\sc LensEnt}.

Another common factor for these two cases is that seeing
conditions during the observations were within the worst quartile of the sample.
This puts in evidence how critical seeing conditions are for ground-based,
weak lensing studies.

\section{Conclusions}

We  analyzed the mass content of galaxy clusters
belonging to a well defined sample of 24 Abell
clusters brighter than
\lx=$5 \times 10^{44}$ \ergs \citep[0.1-2.4 keV][]{XBACS}, spanning
over the redshift range 0.05-0.31, using current techniques of weak-
lensing analysis and homogeneous observing material of subarsecond
image quality. The resulting catalog of mass maps determined for 
22 of these clusters, together with the corresponding light maps, have been
put together in the Appendix. Our main conclusions may be
summarized as follows:

\begin{enumerate}

\item We were able to detect significant weak-lensing signal in 22 out of 24
clusters. This high success rate shows the feasibility of weak-lensing studies
with 8m-class telescopes using service mode observations and relatively short
exposure times. It also indicates that the X-ray luminosity is indeed a good way 
to select massive clusters. Non-detections in A1651 and A1664 
are probably due to a combination of poorer observing conditions with low mass
content. 

\item The center of the mass and light distributions of the clusters
are coincident for $\sim$77\% of the sample (17 out of 22).
 
\item Few clusters present massive substructures, what can be due, in part, 
to small fields (0.4 to 1.8 Mpc on a side) and the relatively low resolution of 
the mass maps. When significant substructures are seen, they are generally
associated with bright cluster members. However, at least in one cluster
(A1451) there seems to exist a substructure
without a clear optical counterpart. 

\item The clusters analyzed here present important departures from spherical
symmetry, as can be verified by the better fits obtained with elliptical
profiles. 

\item We have found, for the first time in an statistically significant sample,
that the dark matter and brightest cluster galaxy major-axes are
strongly aligned: for 62.5\% of the clusters (15 out of 24) 
the difference between their position angle is smaller than 20$\degr$.

\item Most clusters are in or near a state of dynamical equilibrium. This
diagnosis derives from  the agreement between their velocity dispersions 
and the temperature of their ICM, directly measured and/or inferred
from weak-lensing data. Except for A2744, A1451 and  A2163, which
also present evidence of substructures or other complexities,
the other clusters show agreement between these quantities at 
a 1.5 $\sigma$ level. 

\item Clusters in our sample with $T_X > 8$ keV (or $\sigma_v > 1120$ \kms)
show signal of dynamical activity. A2163 and A1451 present large differences
between lensing and dynamical mass estimates and seem to be far from 
equilibrium. In both cases, this conclusion corroborates previous X-ray
analysis.

\item Abell 2744 is the single cluster in this sample that has X-ray
measured temperature more than 1 $\sigma$ below the value inferred by 
weak-lensing. Taking also into account the complex dynamics of this cluster, 
we may explain this discrepancy by assuming that A2744 is a superposition of 
two clusters along the line of sight, near each other or in process of
merging.

\end{enumerate}

Most of these conclusions supports a hierarchical scenario
where massive bodies are formed by the agglomeration of smaller ones, and the
departures of equilibrium described above are indeed an evidence that
some clusters that are at the top of the mass function are still
in the process of active evolution. 

\acknowledgments
This work is based on observations collected at the European Southern
Observatory, Chile (ESO P67.A-0597). ESC and LS acknowledges support by
by Brazilian agencies FAPESP, CNPq and PRONEX, and ESC is grateful to
the Latin American Astronomy Network/UNESCO for funding traveling to Chile.
JPK acknowledges support from  CNRS and Caltech.
ESC also thanks the hospitality of the Department of Astronomy of the
the University of Chile and of the Laboratoire d'Astrophysique of the
the Observatoire Midi-Pyr\'en\'ees. We are also grateful to Gast\~ao Lima
Neto and Ronaldo de Souza for useful discussions and Sarah Bridle
and Phil Marshall for making their softwares, {\sc im2shape} and {\sc LensEnt}
available to us.

\begin{appendix}
\section{Mass and Light Maps \label{figs}}

In this appendix we present figures referent to the weak-lensing analysis. 
In each figure there are 4 panels. In the first 
panel (from left to right) we show the cluster image with superimposed sticks that
represent
the average ellipticities and position angles of the background galaxies. The
size of the stick is proportional to the ellipticity. 
The horizontal stick in the upper right
corner of this panel represents an ellipticity of 0.1. 
We also present an ellipse with the axial ratio and position angle of the best 
fit SIE profile.
In the second panel, we show the mass maps,
that is, the surface density distribution reconstructed from the shear data,
in unities of the critical surface density. The contour are in intervals of
$8\times 10^{11}$ M$_\odot$ Mpc$^{-2}$.
In the third panel the distribution of light coming from cluster members
that belong to the cluster red sequence, in arbitrary unities, smoothed with a
Gaussian similar to the one used to produce the mass map. The countour are in
intervals of $1\times10^{11}$ L$_\odot$ Mpc$^{-2}$ in the V band.
The fourth panel presents the radial shear
profile, where the filled circles represent the average ellipticities projected
tangentially to the direction to the cluster center, and the open squares 
show the ellipticities project in a
direction 45$\degr$ of the tangential one. The solid line represents the 
best-fit SIS profile.

\vskip 1cm 

{\large \bf A version of this paper with all figures can be found at\\
http://www.astro.iag.usp.br/$\sim$eduardo/shear.ps.gz}

\vskip 1cm 

\placefigure{a2744fig}
\placefigure{a0022fig}
\placefigure{a0085fig}
\placefigure{a2811fig}
\placefigure{a1437fig}
\placefigure{a1451fig}
\placefigure{a1553fig}
\placefigure{a1650fig}
\placefigure{a1651fig}
\placefigure{a1664fig}
\placefigure{a2029fig}
\placefigure{a2104fig}
\placefigure{a2163fig}
\placefigure{a2204fig}
\placefigure{a3695fig}
\placefigure{a2739fig}
\placefigure{a2345fig}
\placefigure{a2384fig}
\placefigure{a2426fig}
\placefigure{a3856fig}
\placefigure{a3888fig}
\placefigure{a3984fig}
\placefigure{a2597fig}
\placefigure{a4010fig}

\end{appendix}

\newpage

\figcaption[f8a.eps,f8b.eps,f8c.eps,f8d.eps]   {A2744\label{a2744fig}}
\figcaption[f9a.eps,f9b.eps,f9c.eps,f9d.eps]   {A0022\label{a0022fig}}
\figcaption[f10a.eps,f10b.eps,f10c.eps,f10.eps]{A0085\label{a0085fig}}
\figcaption[f11a.eps,f11b.eps,f11c.eps,f11.eps]{A2811\label{a2811fig}}
\figcaption[f12a.eps,f12b.eps,f12c.eps,f12.eps]{A1437\label{a1437fig}}
\figcaption[f13a.eps,f13b.eps,f13c.eps,f13.eps]{A1451\label{a1451fig}}
\figcaption[f14a.eps,f14b.eps,f14c.eps,f14.eps]{A1553\label{a1553fig}}
\figcaption[f15a.eps,f15b.eps,f15c.eps,f15.eps]{A1650\label{a1650fig}}
\figcaption[f16a.eps,f16b.eps,f16c.eps,f16.eps]{A1651\label{a1651fig}}
\figcaption[f17a.eps,f17b.eps,f17c.eps,f17.eps]{A1664\label{a1664fig}}
\figcaption[f18a.eps,f18b.eps,f18c.eps,f18.eps]{A2029\label{a2029fig}}
\figcaption[f19a.eps,f19b.eps,f19c.eps,f19.eps]{A2104\label{a2104fig}}
\figcaption[f20a.eps,f20b.eps,f20c.eps,f20.eps]{A2163\label{a2163fig}}
\figcaption[f21a.eps,f21b.eps,f21c.eps,f21.eps]{A2204\label{a2204fig}}
\figcaption[f22a.eps,f22b.eps,f22c.eps,f22.eps]{A3695\label{a3695fig}}
\figcaption[f23a.eps,f23b.eps,f23c.eps,f23.eps]{A3739\label{a2739fig}}
\figcaption[f24a.eps,f24b.eps,f24c.eps,f24.eps]{A2345\label{a2345fig}}
\figcaption[f25a.eps,f25b.eps,f25c.eps,f25.eps]{A2384\label{a2384fig}}
\figcaption[f26a.eps,f26b.eps,f26c.eps,f26.eps]{A2426\label{a2426fig}}
\figcaption[f27a.eps,f27b.eps,f27c.eps,f27.eps]{A3856\label{a3856fig}}
\figcaption[f28a.eps,f28b.eps,f28c.eps,f28.eps]{A3888\label{a3888fig}}
\figcaption[f29a.eps,f29b.eps,f29c.eps,f29.eps]{A3984\label{a3984fig}}
\figcaption[f30a.eps,f30b.eps,f30c.eps,f30.eps]{A2597\label{a2597fig}}
\figcaption[f31a.eps,f31b.eps,f31c.eps,f31.eps]{A4010\label{a4010fig}}

\end{document}